\documentclass[12pt,preprint]{aastex}

\newcommand{\htwo}{H\,{\scriptsize\sc{II}} }

\begin{document}

 \title{The Hot and Clumpy Molecular Cocoon Surrounding the Ultracompact \htwo Region G5.89$-$0.39}





 \author{Yu-Nung Su,\altaffilmark{1} Sheng-Yuan Liu,\altaffilmark{1,2} Kuo-Song Wang,\altaffilmark{1} Yi-Hao Chen,\altaffilmark{3} and Huei-Ru Chen\altaffilmark{1,2}}

 \altaffiltext{1}{Institute of Astronomy and Astrophysics, Academia Sinica, P.O. Box 23-141, Taipei 106, Taiwan; ynsu@asiaa.sinica.edu.tw}
 \altaffiltext{2}{Institute of Astronomy and Department of Physics, National Tsing Hua University, Hsinchu, Taiwan}
 \altaffiltext{3}{Department of Physics, National Taiwan University, Taipei, Taiwan}

\begin{abstract}

We present observations of CH$_3$CN (12$-$11) emission at a
resolution of $\sim$2\arcsec~ toward the shell-like ultracompact
\htwo region G5.89$-$0.39 with the Submillimeter Array. The
integrated CH$_3$CN emission reveals dense and hot molecular cocoon
in the periphery of the \htwo region G5.89$-$0.39, with a CH$_3$CN
deficient region roughly centered at G5.89$-$0.39. By analyzing the
CH$_3$CN emission using population diagram analysis, we find, for
the first time, a decreasing temperature structure from 150 to 40
\emph{K} with the projected distance from Feldt's star, which is
thought to be responsible for powering the \htwo region. Our results
further indicate that the majority of the heating energy in the
observed dense gas is supplied by the Feldt's star. From the derived
CH$_3$CN column density profile, we conclude that the dense gas is
not uniformly-distributed but centrally-concentrated, with a
power-law exponent of 5.5 for \emph{r} $\lesssim$ 8000 AU, and 2.0
for 8000 AU $\lesssim$ \emph{r} $\lesssim$ 20000 AU, where \emph{r}
is the distance to Feldt's star. The estimated large power index of
5.5 can be attributed to an enhancement of CH$_3$CN abundance in the
close vicinity of Feldt's star.

\end{abstract}
\keywords{\htwo regions --- ISM: individual (G5.89$-$0.39)
--- ISM: clouds --- stars: formation}

\section{Introduction \label{s-intro}}

G5.89$-$0.39 (hereafter G5.89, also known as W28 A2) is a shell-like
ultracompact (UC) \htwo region with an angular diameter of
$\sim$4\arcsec, presumably powered by a young O-type star
\citep{wc89}. Recently, near-IR observations reported a candidate
for powering the UC \htwo region, an O5-V type star
\citep[][hereafter Feldt's star]{fel03}, although there is a
positional offset of $\sim$1\arcsec~ between the \htwo region center
and the near-IR source. The off-center location probably results
from the migration of Feldt's star \citep{fel03}. The estimated
distance to G5.89 varies from 1.9 kpc to 3.8 kpc \citep[][and
references therein]{hun08}. In this letter, we adopt a distance of 2
kpc, which is favored by most recent studies
\citep[e.g.,][]{wat07,hun08}. Several signposts of high-mass star
formation such as energetic outflows
\citep[e.g.,][]{sol04,wat07,hun08} and maser activities
\citep{fis05,kur04} have been reported toward this region. The
orientation of the reported outflows in various tracers is notably
different, indicating the existence of multiple young stellar
objects in this region \citep[][and reference therein]{hun08}.
Indeed, subarcsecond sub-millimeter observations have identified at
least five dust condensations (denoted as SMA1, SMA2, SMA-N, SMA-E
and SMA-S) \citep{hun08}. With the existence of a young O-type star
still embedded in its natal cloud core, G5.89 provides an ideal
laboratory for studying the physical and chemical conditions of
molecular gas surrounding a newly formed massive star.

In this letter, we present methyl cyanide (CH$_3$CN) observations of
G5.89 at an angular resolution of $\sim$2\arcsec~ with the
Submillimeter Array\footnote{The Submillimeter Array is a joint
project between the Smithsonian Astrophysical Observatory and the
Academia Sinica Institute of Astronomy and Astrophysics, and is
funded by the Smithsonian Institution and the Academia Sinica.}
(SMA). As CH$_3$CN is a symmetric-top molecule, its \emph{K}-ladder
components between adjacent \emph{J} levels are closely spaced in
frequency while dramatically different in excitation energies.
Therefore, an excitation analysis of the CH$_3$CN \emph{K}-ladder
transitions provides an ideal tool for probing (rotational)
temperature of dense molecular gas with small calibration
uncertainties \citep{cum83}. Since radiative transitions are not
allowed across \emph{K}-ladders, the derived rotational temperature
is close to the kinetic temperature if the gas is thermalized. Using
population diagram (PD) analysis \citep{gol99}, we derive the
distribution of the temperatures as well as the filling factor and
column density of CH$_3$CN gas toward G5.89. We then discuss the
implications of the observed temperature, filling factor and
CH$_3$CN column density structures and evaluate the density profile
in this region. A possible variation of CH$_3$CN fractional
abundance is also illustrated.

\section{Observations and Data Reduction}

The observations of CH$_3$CN (12$-$11) in the band of 220 GHz were
carried out with the SMA on 2008 April 17 and May 2. With seven
antennas in the array, the projected baselines ranged from about 9~m
to 120~m (7 to 90 k$\lambda$). The phase center was R.A. =
18$^h$00$^m$30.32$^s$ (J2000) and decl. =
$-$24\degr04\arcmin00.50$\arcsec$ (J2000), and the half-power width
of the SMA primary beam was $\sim$54$\arcsec$. The spectral
resolution was 0.41 MHz, corresponding to a velocity resolution of
$\sim$0.55 km s$^{-1}$. The total available double-sideband
bandwidth was 4 GHz. See \citet{ho04} for more complete
specifications of the SMA. The flux calibrators were Uranus (Apr 17)
and Titan (May 2), and the bandpass calibrators were 3C273 (Apr 17
and May 2) and 3C454.3 (Apr 17). The nearby compact radio sources
1733$-$130 (\emph{S} $\sim$ 2.8 Jy in both Apr 17 and May 2) and
1924$-$292 (\emph{S} $\sim$ 4.4 Jy in Apr 17 and 4.7 Jy in May 2)
served as complex gain calibrators. We calibrated the data using the
MIR software package and made maps using the MIRIAD package. With
uniform weighting, the synthesized beam size was about 3.1$\arcsec$
$\times$ 1.8$\arcsec$ at P.A. of 55.0$^\circ$. The rms noise level
in a single channel of the spectral line images was $\sim$130 mJy
beam$^{-1}$ (equivalently $\sim$0.6 K).

\section{The Spectra and Morphology of the CH$_3$CN Emission \label{result}}

\begin{figure}
  \vspace{-3.cm}
  \epsscale{1.}
  \plotone{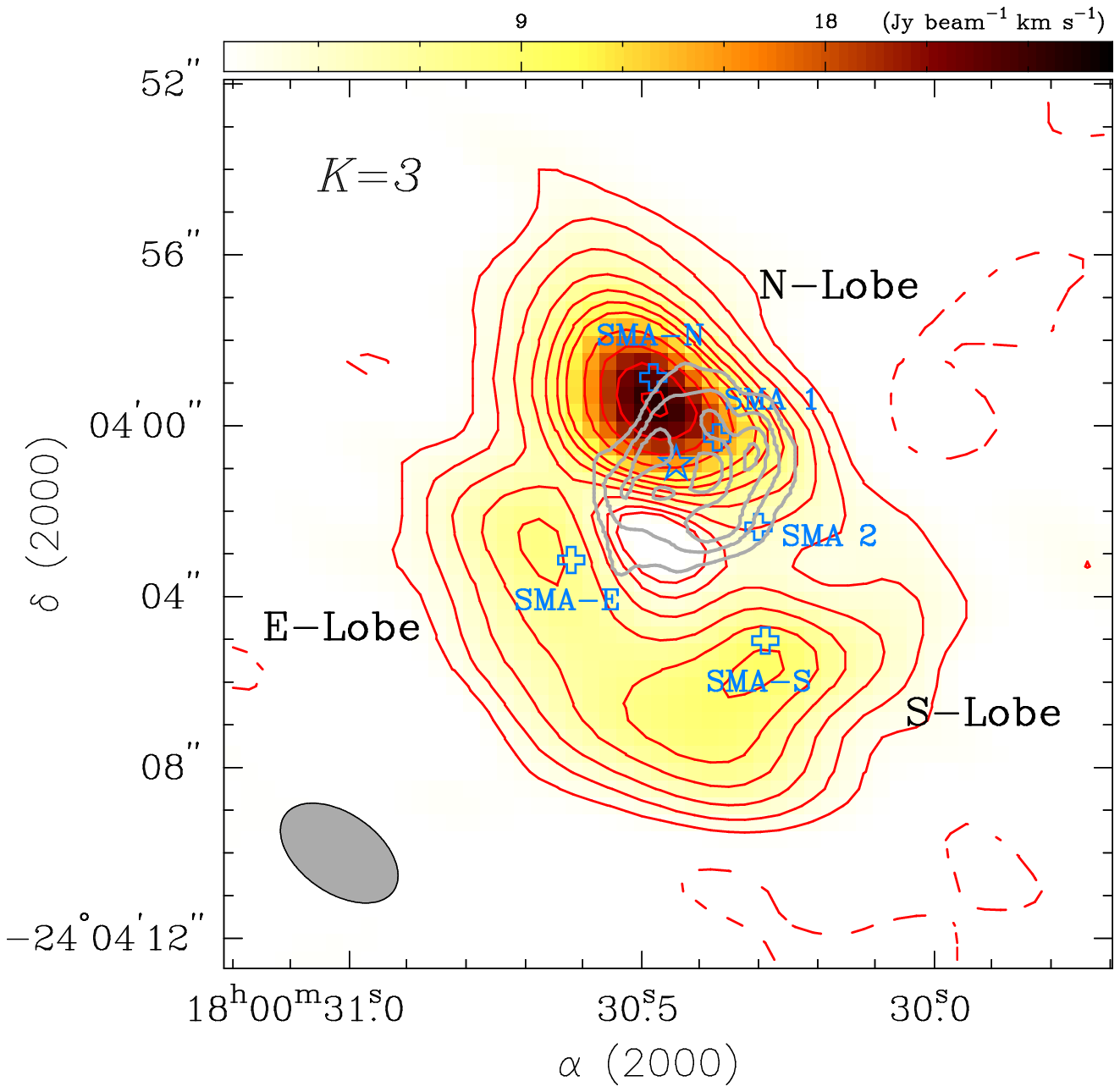}
 \vspace{-4.7cm}
 \caption{The integrated \emph{K} = 3 component of the CH$_3$CN (12$-$11) emission shown in both red contours and color scales.
Contour levels are $-$3, 3, 6, 10, 15, 20, 25, 30, 40, 50, 60, and
70 $\times$ 0.35 Jy beam$^{-1}$ km s$^{-1}$. The gray contours
represent the free-free emission at 2~cm \citep{tan09}. The dark
ellipse at the bottom left denotes the synthesized beam. The crosses
mark the positions of the sub-mm dust condensations reported by
\citet{hun08} and the star marks the position of Feldt's star
\citep{fel03}. \label{fig-ch3cnk3}}
\end{figure}

\begin{figure}
   \vspace{-.5cm}
   \epsscale{0.7}
  \plotone{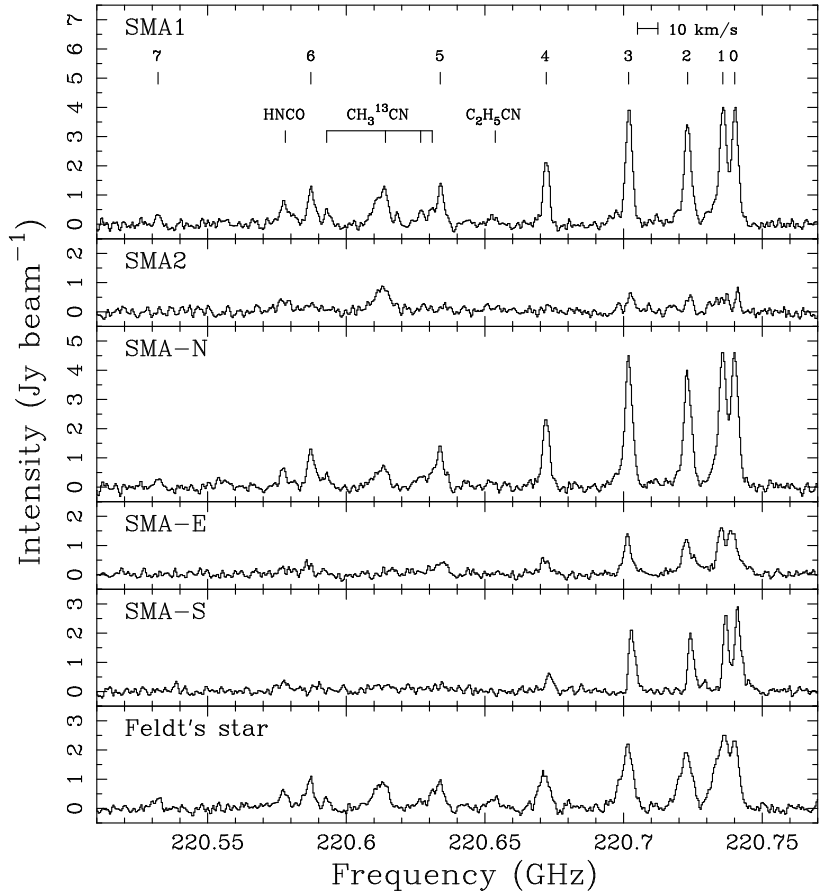}
  \vspace{-2.1cm}
 \caption{The spectra of the CH$_3$CN (12$-$11) lines taken at the positions of the five sub-mm dust condensations identified by \citet{hun08} as well as Feldt's star. \label{fig-spectra}}
\end{figure}

Figure~\ref{fig-ch3cnk3} shows the integrated \emph{K} = 3 component
of the CH$_3$CN (12$-$11) transition. Figure~\ref{fig-spectra} shows
the CH$_3$CN (12$-$11) spectra toward the five sub-mm dust
components identified by \citet{hun08} as well as toward Feldt's
star. The \emph{K} components of the CH$_3$CN (12$-$11) transition
are detected toward G5.89 up to \emph{K} = 7 with an excitation
energy of $\sim$420 K. Comparing the integrated CH$_3$CN flux
measured from our SMA observations with the single-dish results
\citep[e.g.,][]{pan01}, we estimate that approximately 80\% of the
CH$_3$CN emission is recovered by the SMA observations.

The morphologies of the lower \emph{K} (i.e., \emph{K} = 0, 1, and
2) components are similar to that of the \emph{K} = 3 component
shown in Figure \ref{fig-ch3cnk3}. A cavity roughly centered at the
UC \htwo region can be discerned. This cavity is spatially
coincident with the dust-deficient region reported by \citet{hun08},
indicating a deficiency of both molecules and dust grains. With a
synthesized beam of 3.1$\arcsec$ $\times$ 1.8$\arcsec$, the
molecular cocoon seen in the \emph{K} = 3 component is resolved into
at least three peaks, here referred to as N-lobe, E-lobe, and
S-lobe, whose locations agree with the sub-mm dust condensations
SMA-N, SMA-E, and SMA-S, respectively, as shown in Figure
\ref{fig-ch3cnk3}. The higher \emph{K} (i.e., \emph{K} = 5, 6, and
7) components are mainly concentrated in N-lobe (i.e., the vicinity
of SMA-N as well as SMA1) and become much weaker or even undetected
toward E-lobe and S-lobe.

\section{Excitation Analysis of the CH$_3$CN Gas \label{anal}}
\subsection{Methodology \label{anal-pd}}

Given the fairly strong detection of CH$_3$CN transitions toward
G5.89, estimates of gas temperature as a function of position are
feasible. We first performed Gaussian line-profile fitting to
estimate line parameters, including radial velocity, line-width, and
integrated intensity, of each \emph{K} component at each pixel. By
assuming observed transitions to be optically thin and in LTE, we
then calculated
\emph{N}$_\textrm{\scriptsize{u}}^\textrm{\scriptsize{obs}}$, the
CH$_3$CN column density of the upper level of the transition, from
the integrated intensities. For details, see equation (1) in
\citet{tho99}. Since the inferred gas density of $\gtrsim$10$^6$
cm$^{-3}$  (see \S\ref{density}) is at least a factor of 10 higher
than the critical density at the derived gas temperature
\citep{hchen06}, the LTE approximation should be valid.

In the G5.89 region the lower \emph{K} components of the CH$_3$CN
(12$-$11) emission are apparently optically thick, as shown in
Figure \ref{fig-spectra}. Under such conditions, a good estimation
of gas rotational temperature is not achievable by using the
well-known rotation diagram technique \citep{hol82,tun91}. To remedy
this situation, we estimate the gas temperature with the PD analysis
instead. The difference between the PD analysis and the rotational
diagram analysis is that the former includes two additional
correction factors, i.e., the optical depth correction factor
\emph{C$_\tau$} (= $\tau$/(1$-$e$^{-\tau}$)) and the filling factor
\emph{f} (=$\Omega_s$/$\Omega_a$), where $\tau$ is the line opacity,
$\Omega_s$ is the source solid angle, and $\Omega_a$ is the beam
solid angle. Under LTE conditions, the rotational temperatures as
well as the filling factors and column densities of CH$_3$CN can be
determined from the following equation in PD analysis:
\begin{equation}
 ln(\frac{\hat{N}_\textrm{\scriptsize{u}}^\textrm{\scriptsize{obs}}}{g_\textrm{\scriptsize{u}}})~=~ln(\frac{N_\textrm{\scriptsize{tot}}}{Q(T_\textrm{\scriptsize{rot}})})~-~\frac{E_\textrm{\scriptsize{u}}}{kT_\textrm{\scriptsize{rot}}}~+~ln(f)~-~ln(C_\tau)
\end{equation}

where $\hat{N}$$_\textrm{\scriptsize{u}}^\textrm{\scriptsize{obs}}$
is the expected upper-state column density of the target molecule
incorporating the effects of the line opacity (\emph{C$_\tau$}) and
the beam dilution (\emph{f}); \emph{g}$_\textrm{\scriptsize{u}}$ is
the degeneracy of upper state; \emph{N}$_\textrm{\scriptsize{tot}}$
is the total column density of the molecule;
\emph{Q}(\emph{T}$_\textrm{\scriptsize{rot}}$) is the dimensionless
rotational partition function; \emph{T}$_\textrm{\scriptsize{rot}}$
is the rotational temperature; and
\emph{E}$_\textrm{\scriptsize{u}}$ is the upper energy level.

According to equation (1), for a given upper level,
$\hat{N}$$_\textrm{\scriptsize{u}}^\textrm{\scriptsize{obs}}$ can be
evaluated from a set of \emph{N}$_\textrm{\scriptsize{tot}}$,
\emph{T}$_\textrm{\scriptsize{rot}}$, \emph{f}, and \emph{C$_\tau$}.
Since \emph{C$_\tau$} is actually a function of
\emph{N}$_\textrm{\scriptsize{tot}}$ and
\emph{T}$_\textrm{\scriptsize{rot}}$, the independent parameters are
therefore \emph{N}$_\textrm{\scriptsize{tot}}$,
\emph{T}$_\textrm{\scriptsize{rot}}$, and \emph{f} only. We
calculated
$\hat{N}$$_\textrm{\scriptsize{u}}^\textrm{\scriptsize{obs}}$ for
the parameter space of \emph{T}$_\textrm{\scriptsize{rot}}$ =
10$-$300 K, \emph{N}$_\textrm{\scriptsize{tot}}$ in the range of
10$^{13}$$-$10$^{18}$ cm$^{-2}$, and \emph{f} between 0.01 and 1.0.
By comparing
$\hat{N}$$_\textrm{\scriptsize{u}}^\textrm{\scriptsize{obs}}$ with
\emph{N}$_\textrm{\scriptsize{u}}^\textrm{\scriptsize{obs}}$ deduced
from observational results, we performed $\chi^2$ minimization
\begin{equation}
  \chi^2~=~\sum{(\frac{N_\textrm{\scriptsize{u}}^\textrm{\scriptsize{obs}}-\hat{N}_\textrm{\scriptsize{u}}^\textrm{\scriptsize{obs}}}{\delta N_\textrm{\scriptsize{u}}^\textrm{\scriptsize{obs}}})^2}
\end{equation}
to determine for each pixel the physical properties, i.e.,
\emph{T}$_\textrm{\scriptsize{rot}}$,
\emph{N}$_\textrm{\scriptsize{tot}}$, and \emph{f}, from the
selected parameter space.
$\delta{N}_\textrm{\scriptsize{u}}^\textrm{\scriptsize{obs}}$ is the
1-$\sigma$ error of
\emph{N}$_\textrm{\scriptsize{u}}^\textrm{\scriptsize{obs}}$. See
\citet{wan09} for a detailed description of the analysis method.

\subsection{Temperature \& Illumination of Molecular Gas \label{dis-temp}}

As shown in Figure \ref{fig-results}\emph{a}, we have for the first
time resolved the temperature structure of the neutral dense gas in
this region with the PD analysis. The inferred CH$_3$CN gas
temperatures range from 40 \emph{K} to 150 \emph{K}, in agreement
with the single-dish results of 56$-$75 \emph{K}
\citep{tho99,pan01,pur06}. Among the three CH$_3$CN lobes, N-lobe is
the hottest, E-lobe next, and S-lobe is the coldest. In particular,
the CH$_3$CN gas in the close vicinity of Feldt's star has the
highest temperature within the whole region.

To further explore the role of Feldt's star in illuminating its
surrounding gas, we plot the deduced gas temperature
\emph{T}$_\textrm{\scriptsize{rot}}$ versus the projected distance,
\emph{d}, to Feldt's star. Indeed, a good correlation between the
temperature and the projected distance is revealed in Figure
\ref{fig-tempvspdist}\emph{a}. In contrast, the correlation between
\emph{T}$_\textrm{\scriptsize{rot}}$ and \emph{d} becomes much more
vague when the fiducial center is shifted about 2$\arcsec$ in any
direction away from Feldt's star. Together with the sufficient
luminosity budget of the O5-type Feldt's star and the non-detection
of other candidate \htwo region powering stars within 2$\arcsec$
from Feldt's star, we conclude that the majority of the energy for
heating the dense gas enshrouding the UC \htwo region G5.89 comes
from Feldt's star.

A power-law fit to the data indicates that the gas temperature
scales as \emph{T}$_\textrm{\scriptsize{rot}}$ $\propto$
\emph{d}$^{-0.40}$. Note that the deduced temperature
\emph{T}$_\textrm{\scriptsize{rot}}$(\emph{d}) is actually a
representative or averaged temperature for the gas along the line of
sight. We suggest, however, that with a highly
centrally-concentrated density distribution as shown in
\S\ref{density}, the deduced temperature at small \emph{d} is
heavily weighted by high density, hence this represents the gas
temperature at small \emph{r}, where \emph{r} is the distance to
Feldt's star. Therefore the temperature profile
\emph{T}$_\textrm{\scriptsize{rot}}$(\emph{d}) can be viewed as
\emph{T}$_\textrm{\scriptsize{rot}}$(\emph{r}). The temperature
profile of \emph{T} $\propto$ \emph{r}$^{-0.40}$ is shallower than
the profile of \emph{T} $\propto$ \emph{r}$^{-3/4}$ measured toward
hot-molecular core sources \citep[e.g.,][]{ces98,bel05}. This
disparity can be attributed to different dust grain properties
and/or variant evolutionary stages. The temperature structure
provides clues for the dust opacity index $\beta$. Theoretical
models found that, for example, the temperature structure of uniform
dust shells surrounding a newly formed O star can be described by
\emph{T$_{dust}$} $\propto$ \emph{r}$^{-\alpha}$, where $\alpha$
$\approx$ 2/($\beta$+4) \citep{wol94}. Modifications of the models
with a centrally-concentrated density profile, however, are required
to match the case of G5.89.

\subsection{Filling Factors \& Column Density Profiles
\label{dis-fillingf}} 
\begin{figure}
  \vspace{-5.cm}
  \epsscale{1.03}
  \plotone{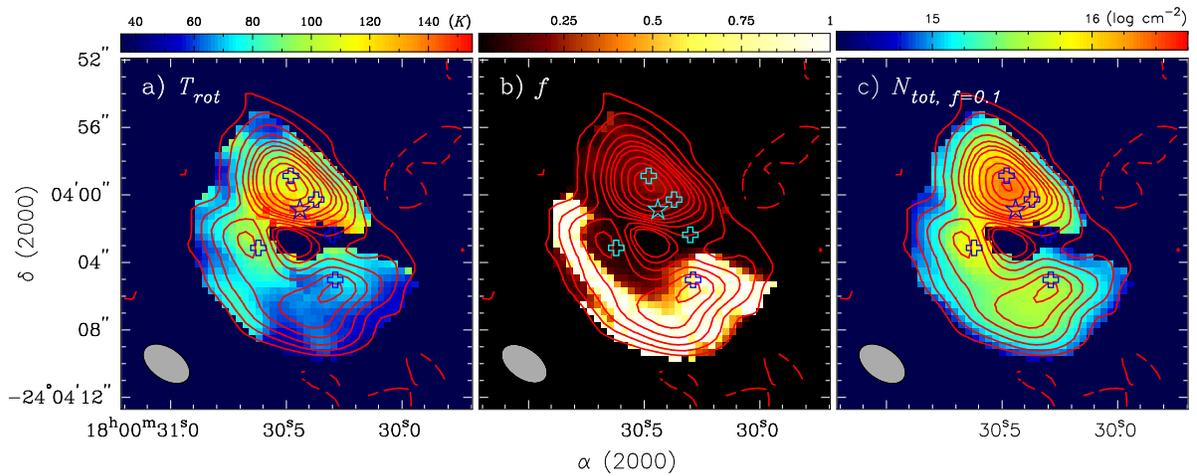}
 \vspace{-9.cm}
 \caption{The inferred rotational temperature (panel \emph{a}), filling
factor (panel \emph{b}), and CH$_3$CN column density (panel
\emph{c}) in color scales overlaid with the contours of the
integrated CH$_3$CN (12$-$11) \emph{K} = 3 component. The CH$_3$CN
column density plotted in panel \emph{c},
\emph{N}$_\textrm{\scriptsize{tot}},_{ f=0.1}$, is the CH$_3$CN
column density scaled by a filling factor of 0.1. Caption as in
Figure \ref{fig-ch3cnk3}. \label{fig-results}}
\end{figure}

Figures~\ref{fig-results}\emph{b} and
\ref{fig-results}\emph{c} show the derived filling factor and
CH$_3$CN column density, respectively.  The CH$_3$CN column density
plotted in Figure \ref{fig-results}\emph{c},
\emph{N}$_\textrm{\scriptsize{tot}},_{ f=0.1}$, is the CH$_3$CN
column density scaled by a filling factor of 0.1 (see below).
Figure~\ref{fig-tempvspdist}\emph{b} shows the derived filling
factor versus the projected distance to Feldt's star, and it clearly
demonstrates that the molecular gas in the close vicinity of Feldt's
star (i.e., \emph{d} $\lesssim$6000 AU, mainly associated with
N-lobe) has small filling factors ranging from 0.02 to 0.24,
suggestive of a clumpy medium. The small filling factors can also be
readily recognized from the low brightness temperatures
($\lesssim$20 K) of the hot ($\sim$100$-$150 \emph{K}) but optically
thick low-\emph{K} components. For the gas further out, the filling
factors become noticeably scattered with not only large values of
0.8$-$1 for the majority but also large uncertainties. Nevertheless,
such large uncertainties in fact imply all the observed transitions
to be optically thin. In such case, the filling factor, \emph{f},
and the total column density, \emph{N}$_\textrm{\scriptsize{tot}}$,
become degenerate in the calculation and hence cannot be well
determined.

In short, the CH$_3$CN gas toward G5.89 can be coherently considered
as clumpy, with \emph{f} $\sim$ 0.1.$-$0.2. As shown in Figure
\ref{fig-tempvspdist}\emph{c}, we thus plot the CH$_3$CN column
density scaled by a filling factor of 0.1,
\emph{N}$_\textrm{\scriptsize{tot}},_{ f=0.1}$, versus the projected
distance to Feldt's star. Obviously
\emph{N}$_\textrm{\scriptsize{tot}},_{ f=0.1}$ decreases with
increasing projected distance \emph{d}, and the ratio between the
inner (\emph{d} $\sim$ 4000 AU) and outer (\emph{d} $\sim$ 18000 AU)
region is more than a factor of 10. Note that if
\emph{N}$_\textrm{\scriptsize{tot}}$ instead of
\emph{N}$_\textrm{\scriptsize{tot}},_{ f=0.1}$ is plotted, this
ratio will be even larger because the inferred \emph{f} is larger
for gas further out.

\section{Density Structure \label{density}}
\begin{figure}
   \epsscale{0.7}
  \plotone{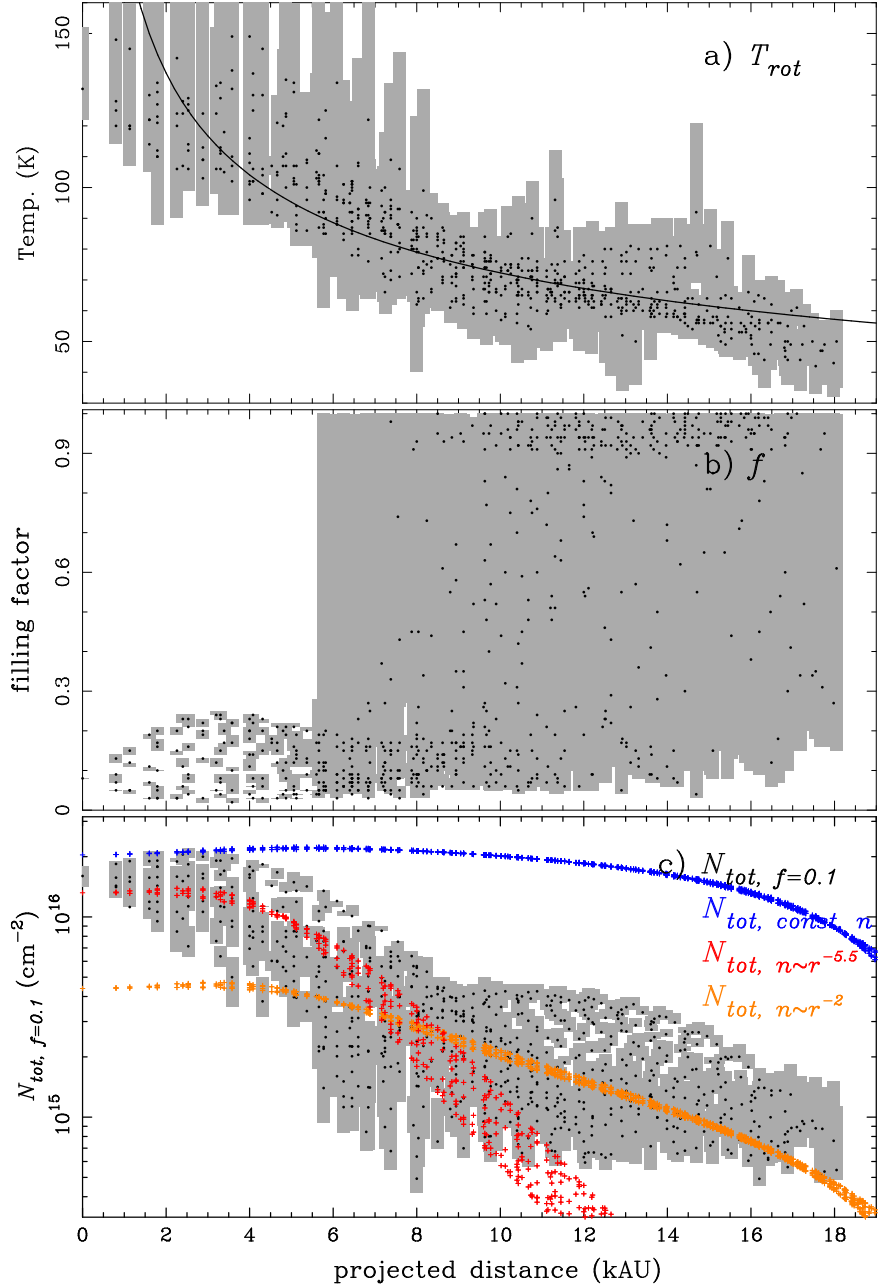}
  \caption{The inferred rotational temperature (panel \emph{a}), filling
factor (panel \emph{b}), and CH$_3$CN column density (panel
\emph{c}) versus the projected distance, \emph{d}, to the Feldt's
star.   In each panel, the overlaid greyscales represent the
1-$\sigma$ error in y-axis of the plotted data. In panel \emph{a},
the best power-law fit, \emph{T}$_\textrm{\scriptsize{rot}}$
$\propto$ \emph{d}$^{-0.40}$, is also plotted. In panel \emph{c},
the blue, orange, and red crosses represent the modeled CH$_3$CN
column densities with the spherical density distributions of
power-law indices, \emph{p}, of 0, 2, and 5.5. See \S\ref{density}
for the details. \label{fig-tempvspdist}}
\end{figure}

The deduced CH$_3$CN column density profile reflects the density
structure of the molecular gas. For simplicity, we assume a
spherical cloud core with a power-law density profile,
\emph{n}(\emph{r}) $\sim$ \emph{r}$^{-p}$, and a constant CH$_3$CN
fractional abundance \emph{X}(CH$_3$CN) of 1 $\times$ 10$^{-9}$,
similar to that found in other massive star forming regions
\citep[][]{wil94,zha98}. We further adopt inner and outer boundaries
at 4000 and 20000 AU, respectively, based on the size of the UC
\htwo region and the CH$_3$CN emission. We then convolve the column
density with a two-dimensional Gaussian beam of 3.1$\arcsec$
$\times$ 1.8$\arcsec$, equivalent to our SMA beam. As shown in
Figure \ref{fig-tempvspdist}\emph{c}, we examine three model column
density profiles calculated with power-law indices \emph{p} of 0, 2,
and 5.5. The adopted values for density at the inner edge are
3.5$\times$10$^{7}$, 3.5$\times$10$^{7}$, and 3.5$\times$10$^{8}$
cm$^{-3}$, respectively. Obviously, the modeled column density
profile of \emph{p}=0 (i.e., constant density) is too flat to agree
with the observational results (i.e.,
\emph{N}$_\textrm{\scriptsize{tot}},_{ f=0.1}$ profile). Since the
\emph{N}$_\textrm{\scriptsize{tot}}$ profile would be even steeper
than \emph{N}$_\textrm{\scriptsize{tot}},_{ f=0.1}$, the discrepancy
between the profiles of observations and constant density can not be
attributed to the adopted constant filling factor of 0.1. The
observed \emph{N}$_\textrm{\scriptsize{tot}},_{ f=0.1}$ profile can
be well represented by the results modeled from a broken power law
density distribution, with \emph{p} = 5.5 for \emph{r} $\lesssim$
8000 AU and \emph{p} = 2.0 for 8000 AU $\lesssim$ \emph{r}
$\lesssim$ 20000 AU. Note that for the \emph{p} = 2 profile, the gas
density at \emph{r} = 20000 AU is 1.4 $\times$ 10$^6$ cm$^{-3}$,
more than sufficient to thermalize the observed CH$_3$CN
transitions. We therefore conclude that a centrally-concentrated
density distribution is required to interpret the observational
column density profile.

Observations of massive star-forming cores indicated that the
density power-law indices range from 1 to 2, while theoretical
models also predicted similar power-law indices \citep[e.g.,][and
reference therein]{mue02}. The density profile of \emph{n} $\sim$
\emph{r}$^{-5.5}$ is significantly steeper than typical cases.
Feedback processes such as the expansion of the \htwo region may
play a role to steepen the density profile. Since the observed
CH$_3$CN column density depends on the gas density and CH$_3$CN
abundance, \emph{X}(CH$_3$CN), the steep CH$_3$CN column density
profile may also be a result of an enhancement of CH$_3$CN gas due
to the evaporation of grain mantles in the inner region, similar to
the cases of abundance jump reported by \citet{tak00}.

Finally, we note that the temperatures as well as column densities
are estimated simultaneously under the assumption of one isothermal
gas component along the line of sight. While the deduced temperature
and density structures are in fact not uniform, they form a
self-consistent set of solutions. That is, the highly
centrally-concentrated CH$_3$CN density profile supports the
temperature profile, which in term validates the density profile
itself. The derived ``projected'' profile therefore reflects the
actual radial-dependent structure. More rigorous radiative transfer
calculation that can account for all the involved parameters such as
temperature, density, filling factors, and abundance, as well as
observations with even higher angular resolutions will be desired to
further refine the results.

\acknowledgments We thank all SMA staff for their help during these
observations. We thank J. Karr for reading the manuscript. S.-Y. L.,
Y.-N. S., and H.-R. C. thank the National Science Council of Taiwan
for support this work through grants NSC 97-2112-M-001-006-MY2 and
NSC 97-2112-M-007-006-MY3.

\clearpage

\clearpage

\clearpage

\clearpage

\end{document}